\documentclass{revtex4}

\usepackage{epsfig}
\usepackage{float}
\usepackage{graphicx}    
\usepackage{dcolumn}     
\usepackage{bm}          
\usepackage{subfigure}   
\usepackage{citesort} 

\graphicspath{{Figures/}}

\begin{document}

\title{Keynesian Economics After All}
\author{A.\ Johansen}
\email{agung.urum@gmailcom}
\affiliation{Teglg\aa rdsvej 119, DK-3050 Humleb\ae k, Denmark}
\author{I.\ Simonsen}
\email{Ingve.Simonsen@phys.ntnu.no}
\affiliation{Department of Physics, 
       Norwegian University of Science and Technology (NTNU), 
       NO-7491 Trondheim,
       Norway} 

\date{\today}

 \begin{abstract}
   It is demonstrated that the US economy has on the long-term in
   reality been governed by the Keynesian approach to economics
   independent of the current official economical policy.  This is
   done by calculating the two-point correlation function between the
   fluctuations of the DJIA and the US public debt. We find that the
   origin of this condition is mainly related to the wars that the USA
   has fought during the time period investigated.  Wars mean a large
   influx of public money into the economy, thus as a consequence
   creating a significant economical upturn in the DJIA. A reason for
   this straight-cut result of our analysis, is that very few wars
   have been fought on US-territory and those that have, were in the
   18th century, when the partial destruction of cities, factories,
   railways and so on, was more limited and with less effect on the
   over-all economy.
 \end{abstract}

 \maketitle


What drives prices? This question has been debated and studied for
centuries without any definite answer emerging. Especially in times
of financial turmoil, where stocks in many cases are trading far
below the so-called fundamental values, is this question important.

Besides the economical framework of Karl Marx, which we will not
consider here since the topic of the present paper is the US markets,
two main economical paradigms~\footnote{We call them ``paradigms'' and
  not ``theories'', as most economics are axiomatic.} exist. One is
due to the work of Adam Smith~\cite{AS} and advocates (more or less)
completely free market conditions. The other is due to the work of
John M. Keynes~\cite{JK1,JK2} and, independently, Gunnar
Myrdal~\cite{GM1,GM2,GM3,GM4}.
Below we will briefly review and comment on these two paradigms. As
the former has been the ruling paradigm in the western world for more
than three decades, as well as in previous periods, {\it e.g.}, before
and after the infamous crash of 1929 until the implementation of the
``New Deal'', it can certainly be argued that it is responsible for
the current financial turmoil.  In support of this stand the quick
shift in financial policies of the central banks all in most western
economies, as well as others (China, Japan,{\it etc}.), can without
any doubt be fitted into the framework of Keynesian economics.

\bigskip

Adam Smith~(1723--1790) believed that the markets were in constant
equilibrium as the intervention of an ``invisible hand'', {\it i.e.}
the market forces, would eliminate any inequality in demand and supply
by adjusting the price of the commodity in question. This idea has
been re-formulated in the framework of purely selfish hard-working
rational traders with complete knowledge of all available information
whose continuing effort more or less instantaneously removes any
imbalance in prices (arbitrage opportunities) due to past differences
in the expectations of traders. This model has been coined the
``Efficient Market Hypothesis''~(EMH)~\cite{EMH1,EMH2}.

The EMH states that current prices reflect all available information
about the priced commodity; in other words, all available information
is at any given instant already priced in by the market and any change
in prices can only be due to the revelation of new information, {\it
  i.e.}, there is no ``free lunch''.

On the macro-economical level, the EMH has meant that only monetary
policies --- changes in the interest rates of the central bank and the
money balance~\footnote{``Money balance'' refers essentially to the
  supply of money. The central banks can affect the money balance by
  buying or selling treasury securities. Secondly, the discount rate
  can be changed. And finally, the Federal Reserve can adjust the
  reserve requirement of private banks.} --- are ``allowed''. The
``rest'' is best dealt with through the forces of the markets.  Since
the 1980s and until the recent economical crises the EMH has been the
dominant economical paradigm and has led to a massive deregulation of
the financial markets, global trade and floating exchange rates.
Quite ironically from today's perspective, this kind of {\it laissez
  faire} economics in fact led to the Great Depression and not, as
commonly believed, the crash of 1929~\footnote{In November 1920,
  Warren G. Harding was elected president followed by presidents
  Coolidge and Hover, leading to 12 consecutive years of Republican
  control of the White House and strongly pro-business government
  policy. In practice this meant as little government influence as
  possible, since the administrations of Presidents Harding,
  Coolidge, and Hoover all believed the rules of Adam Smith still
  applied.}.

As EMH only addresses the relation between prices and available
information, it can only be tested jointly with some asset-pricing
model.  This introduces an ambiguity in how to interpret anomalous
behaviour of returns: Is the market inefficient to some extent or do
we have an inaccurate price model?  This to some extent turns the
entire question into a matter of ``religion''~\cite{thesis}.

Another, more general problem, with standard models of financial
markets is the following: If one assumes, as these models do, that the
market participants are purely selfish individuals that optimize
their own utility function through fixed time contracts with other
nominally identical individuals, then, despite the achievement of
mutual benefits, the term ``contract'' would not be defined in a
general context. This is so because a general and lasting definition
of the term ``contract'' requires long-term (meaning much longer than
an ordinary human time span) institutions, which can only be upheld
through non-selfish behavior. For example regulatory institutions have
been built during centuries and hence cannot be put to work within a
framework of purely selfish individuals who's time horizon can't
possibly be longer than their own life span. This means that important
psychological and social factors must exist to counter a purely
selfish behaviour. Hence, as most models of the financial markets are
based on EMH, they lack two very essential ingredients, which one
usually denote by the general word ``culture'' and
``psychology''~\cite{PRE}.

\medskip

The second economical paradigm is due to John M. Keynes (1883--1946)
and Gunnar Myrdal (1898--1987)~\cite{JK1,JK2,GM1,GM2,GM3,GM4}. The
basis of Keynes economical work is that at times, market forces will
push the economy so far from equilibrium that it is necessary for
governments and central banks to take rigorous financial actions such
as increased government spending and tighter regulation of the
markets. In short, using the characterization attributed to Keynes;
``The markets can stay irrational longer than you can stay solvent''.

The work of Keynes was highly popular among the governments of the
western world after the Second World War~(WW2) and until the 1980s
partly due to the success of Roosevelt's ``New Deal'' policy designed
to deal with the financial disaster initiating the Great Depression of
the 1930s.  The most prominent post-WW2 example is the Marshall aid,
but also the establishment of socialized benefits in the 1960s, such
as unemployment insurance and state subsidized education and medical
care, are noteworthy examples of Keynesian economics. Also, the
concept of devaluation or revaluation of a country's currency is a
prime example of Keynesian economics.

It should be mentioned that the work of Keynes has been criticized by
many economists from a theoretical standpoint as it lacks a firm
theoretical basis, {\it i.e.}, it is more pragmatic than
axiomatic. This discussion is beyond the scope of the present
paper. However, the Swedish economist Gunnar Myrdal has in part
provided the work of Keynes with a more firm framework, besides
incorporating sociological elements into
economics~\cite{GM1,GM2,GM3,GM4}.  For this contribution G.~Myrdal was
in 1974 awarded Sveriges Riksbank Prize in Economic Sciences in Memory
of Alfred Nobel (unofficially the Nobel Price in
Economics)~\footnote{As Alfred Nobel didn't believe economy to be a
  science, he did not establish a prize in this field. In 1968, the
  Swedish Central Bank established a pseudo-Nobel Prize in
  economy. Recently, there has been some controversy around the name
  of this prize, as some members of the Nobel family want to stop the
  use of the name ``Nobel'' for this prize.}.

\medskip

\begin{figure}[t]
  \centering
  \includegraphics*[height=9.5cm,width=13.5cm]{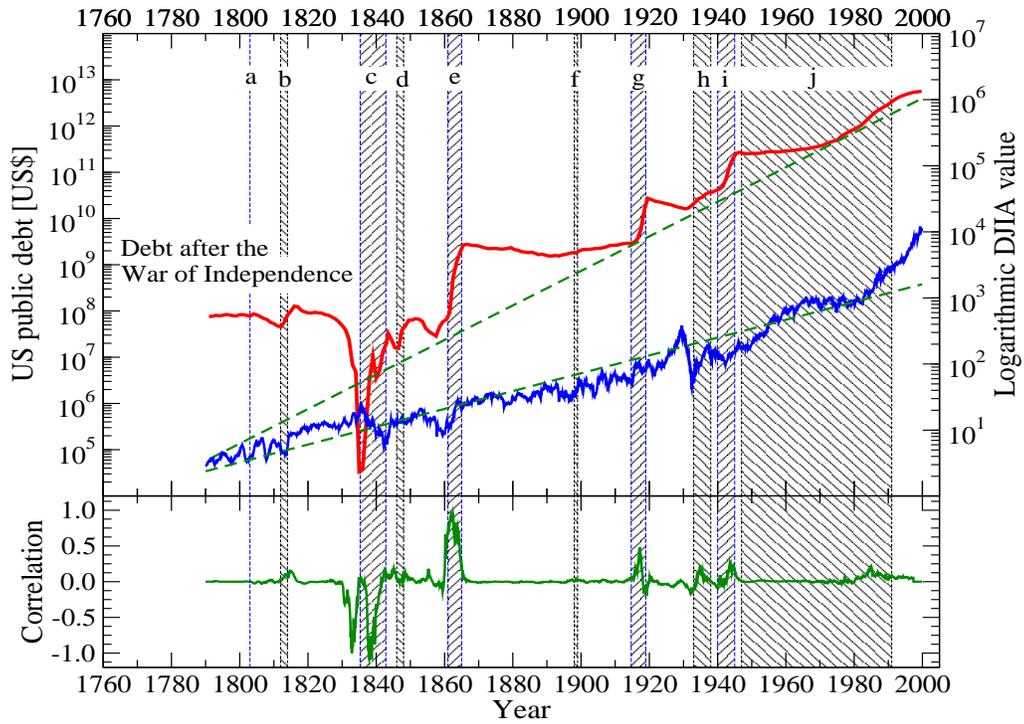}
  \caption{(top panel) The historical US public debt~(red curve) and the
    value of the DJIA~(blue curve) from 1791 till 2000. 
    The two dashed diagonal lines represent exponential functions corresponding to
    an average growth rate of about $8.6\%$ (for the debt) and $2.5\%$
    for the DJIA index. 
    Some historic events are marked by dashed areas in the figure: (a)
    the 1812 war~(1812--1814); (c) the second war with the Seminole
    Indians~(1835--1842); (d) The Mexican-American War (1846--1848);
    (e) The Civil War~(1861--1865); (f) The Spanish American
    War~(1898); (g) The First World War~(1914--1918); (h) the ``New
    Deal'' policy~(1933–-1938); (i) The Second World War (1940--1945)
    (j) The Cold War~(1947--1991). (bottom panel) The time dependent $5$ year gliding
    two-point correlation function, $C_{\Delta t}(t)$, between the US public
    debt and the value of the DJIA index.}
  \label{publicdebt} 
\end{figure}

We will now show that the primary long-term factor, {\it i.e.} over
decades, driving the US economy during the entire history of the USA
has been of the Keynesian type independent of what the ruling
economical paradigm has been. To this end we present in
Fig.~\ref{publicdebt} the historical public debt of the US together
with a labeling of the most significant historic events, mostly
military conflicts, involving the US during the same period.  Notice
that the Korean War~(1950--1953) with its formal UN-forces did not
represent any significant increase in US public debt. However, the
so-called Cold War~(1947--1991) between the USA and the USSR most
certainly did.  It is clear that the over-all rise in the public debt
is exponentially driven by large ``bumps'' signifying rapid growth in
public debt. It is striking that, on a qualitative level, the origin
of these large growth periods in US public debt is mainly related to
wars with two major exceptions; the purchase of the Louisiana
Territory from Napoleon in 1803 on the unauthorized initiative of the
US-ambassador to France, and the Keynesian (in the meaning massive
public investments) ``New Deal'' policy of Roosevelt in the
1930s. Note, that the use of a logarithmic scale in
Fig.~\ref{publicdebt} is the reason why the ``bumps'' associated with
the purchase of the Louisiana Territory and the Spanish War of 1898 do
not ``stand out''.

If one compares the US public debt to the behaviour of the US stock
market here quantified by the DJIA (Fig.~\ref{publicdebt}), one
clearly sees that on a qualitative level the rises in the public debt,
due to wars and New Deal, are followed by steep rises in the stock
market with one big exception, namely the bubble of the 1920s. In
hindsight, this may not be so surprising, since increases in public
debt normally means large public spending which funnels large amounts
of money into the private sector.  This relation can be made
quantitative by defining a (gliding) two-point correlation function
between the US public debt, $d(t)$, and the DJIA index, $p(t)$, in the
following way
\begin{eqnarray}
  C_{\Delta t}(t) &=& 
    \frac{ 
          \Big<
             \left[ d(t) -\left<d(t)\right>_{\Delta t}\right]
             \left[ p(t) -\left<p(t)\right>_{\Delta t}\right]
          \Big>_{\Delta t}
         }{
           \sigma_{d}(t;\Delta t)\, \sigma_{p}(t;\Delta t)
         },
\end{eqnarray}
where $\left<s(t)\right>_{\Delta t}$ and $\sigma_s(t,\Delta t) = \big<
\left[ s(t) -\left<s(t)\right>_{\Delta t}\right]^2 \big>_{\Delta
  t}^{\frac{1}{2}}$ denote the average and root-mean-square-value,
respectively, of the time series $s(t)$ over a time window of length
$\Delta t$ that is centered at $t$. In this way, and with $\Delta
t=5$~years, we obtained the correlation function shown in the lower
panel of Fig.~\ref{publicdebt}.

We note that the annual average growth rate of the US public debt is
about $8.6\%$ (upper green dashed line in Fig.~\ref{publicdebt}). This
figure should be compared to the average annual growth rate of the
DJIA that until the 1950s was about $2.5\%$. This discrepancy
basically explains the exponential growth in the US public debt. What
is surprising, however, is that most time periods with peace exhibit a
significant decline in the public debt, most notably the period after
the war of 1812 until the second war with the Seminole Indians
(1835--1842)\footnote{``Surprisingly'', the Seminole Indians did not
  want to move from Florida to Idaho.}, as well as modest growth in
the DJIA. In the first half of the 19th century, the US public debt
dropped to a meager US\$~33\,733 in 1835 from US\$~75\,463\,477 in
1791~\footnote{In 1790, the Federal Government declared that it was
  redeeming the Scrip Money that was issued during the Revolutionary
  War in the amount of US\$~80\,000\,000.}. Hence, with respect to
long-term growth in the stock market, public spending, especially in
the case of war, has played a {\it very} significant role in the
long-term growth of the US economy and therefore of the DJIA.

\bigskip

In conclusion, we have demonstrated, that the main long-term driving
force of the US economy through the entire history of the USA has
been massive public investments in the private sector, financed by
significant increases in public debt, during times of war.

\acknowledgments
I.S. acknowledges the support from the European Union COST Action
MP0801 (``Physics of Competition and Conflicts'').



%

\end{document}